\documentclass[a4paper]{jpconf}
\usepackage{graphicx}
\begin{document}
\title{Nanodiamond photocathodes for MPGD-based single photon detectors at future EIC}

\author[add-trieste]{C.~Chatterjee$^{2}$, G. Cicala$^4$, A.~Cicuttin$^{5}$ P.~Ciliberti$^2$, M.L.~Crespo$^5$, S.~Dalla~Torre$^1$, S.~Dasgupta$^1$, M.~Gregori$^1$, S.~Levorato$^1$,   G.~Menon$^1$, F.~Tessarotto$^1$, Triloki$^{5}$, A.Valentini$^3$,L. Velardi$^4$,Y.~X.~Zhao$^1$}

\address{$^1$ INFN Trieste, Trieste, Italy}
\address{$^2$ University of Trieste and INFN Trieste, Trieste, Italy}
\address{$^3$ University Aldo Moro of Bari and INFN Bari, Bari, Italy}
\address{$^4$ CNR-ISTP and INFN  Bari, Bari, Italy }
\address{$^5$ Abdus Salam ICTP, Trieste, Italy and INFN Trieste, Trieste, Italy }

\ead{Triloki@ts.infn.it}

\begin{abstract}
The design of a Ring Imaging CHerenkov (RICH) detector for the identification of high momentum particles at the future Electron Ion Collider (EIC) is extremely challenging by using current technology. Compact collider setups impose to construct RICH with short radiator length, hence limiting the number of generated photons. The number of detected photons  can be increased by selecting the far UV region. As standard fused-silica windows is opaque below 165 nm, a windowless RICH can be a possible approach. CsI is widely used photocathode (PC) for photon detection in the far UV range. Due to its hygroscopic nature it is very delicate to handle. In addition, its Quantum Efficiency (QE) degrades in high intensity ion fluxes. These are the key reasons to quest for novel PC with sensitivity in the far UV region. Recent development of layers of hydrogenated nanodiamond powders as an alternative PC material and their performance, when coupled to the THick Gaseous Electron Multipliers (THGEM)-based detectors, are the objects of an ongoing R\&D. We report here some preliminary results on the initial phase of these studies.
\end{abstract}

\section{Introduction}
The Electron Ion Collider (EIC)~\cite{EIC} will answer many fundamental questions of nuclear and particle physics. In particular, those related to the origin of nucleon mass, spin, and the properties of dense gluon systems. The success of this future EIC laboratory will largely depend on efficient hadron Particle IDentification (PID) at high momenta, namely above $6$-$8~GeV/c$. A gaseous Ring Imaging CHerenkov (RICH) is therefore an obvious choice for hadron PID. However, the number of Cherenkov photons generated in a light radiator is limited. At present, these number of photons is recovered by using long radiators. The compact design of the experimental setup at an EIC collider is a limitation for this approach.  In the far UV spectral region ($\sim120$~nm), the number of generated Cherenkov photon increases, according to the Frank-Tamm distribution.  This direct the detection of photons in the very far UV range.  The standard fused-silica windows are opaque for  wavelengths below 165~nm. Therefore, a windowless RICH is a
potential option. The windowless concept also points to the use of gaseous photon detectors operated with the radiator gas itself~\cite{windowless-RICH}.
\par 

For the detection of single photon in  Cherenkov imaging counters, the MicroPattern Gaseous Detector (MPGD)-based Photon Detectors (PD) have recently been demonstrated as effective devices ~\cite{PM18}. These detectors have a hybrid structure, where two layers of THGEMs are followed by a MICROMEGAS stage; the top layer of the first THGEM is coated with CsI, which acts as a reflective PC.

\par 
Since a last few decades CsI is an obvious PC in the far UV domain. Although CsI has a high photo conversion efficiency in the far UV regions, it has some concerning limitations. Due to its hygroscopic nature, which causes a degrade in QE when it exposed to humid atmosphere, its handling is delicate. Moreover, its QE degradation is also caused by intense ion  bombardment. Ions generated during the multiplication processes in gaseous detectors cause the degradation of QE~\cite{NIMA_574_2007_28}, clearly observed after an integrated charge of the order of $(1~mC/cm^{2}) $ .

The search for an alternative UV sensitive photocathode  without these limitations has therefore been a prime goal for the R\&D of the future EIC RICH. In the present article, we have shown the preliminary results on  nanodimond (ND) and hydrogenated nanodiamond (H-ND) coated THGEM detectors.


\section{Nanodiamond particles as an alternative to CsI}
The high QE value of CsI photocathode in the UV spectral range makes it the mostly used photoconverter for the UV sensitive devices. This high QE value is related to its low electron affinity ($0.1~eV$) and wide band gap ($6.2~eV$). 
The diamond has a band gap of $5.5~eV$ and low electron affinity of $0.35$-$0.50~eV$ and it also exhibits chemical inertness, radiation hardness and good thermal conductivity.  Additionally, hydrogenation of diamond surface lowers the electron affinity down to -$1.27~eV$. The negative electron affinity allows an efficient escape into vacuum of the generated photoelectrons without an energy barrier at the surface~\cite{NDRep-1}. Recently, it has been reported~\cite{NDRep-1, NDreport}  a novel procedure developed in Bari, where hydrogenation of nanodiamond powder allows to achieve strikingly high and stable QE. A comparison of CsI and ND, can be extracted from the literature ~\cite{NDRep-1, NIMA_502_2003_76}.  

\section{Pre-characterization and coating}
\subsection{Characterization before Coating}

\par For the starting exercise of our R\&D studies, we coated five THGEMs. THGEMs are electron multipliers of the MPGD technology. They are actually derived from GEM architecture with thicker dielectric material, a PCB in our case, in between two conductive electrode layers by Cu, with Ni-Au coating.

\begin{figure}
\begin{minipage}[c]{1\textwidth}
    \includegraphics[width=\textwidth]{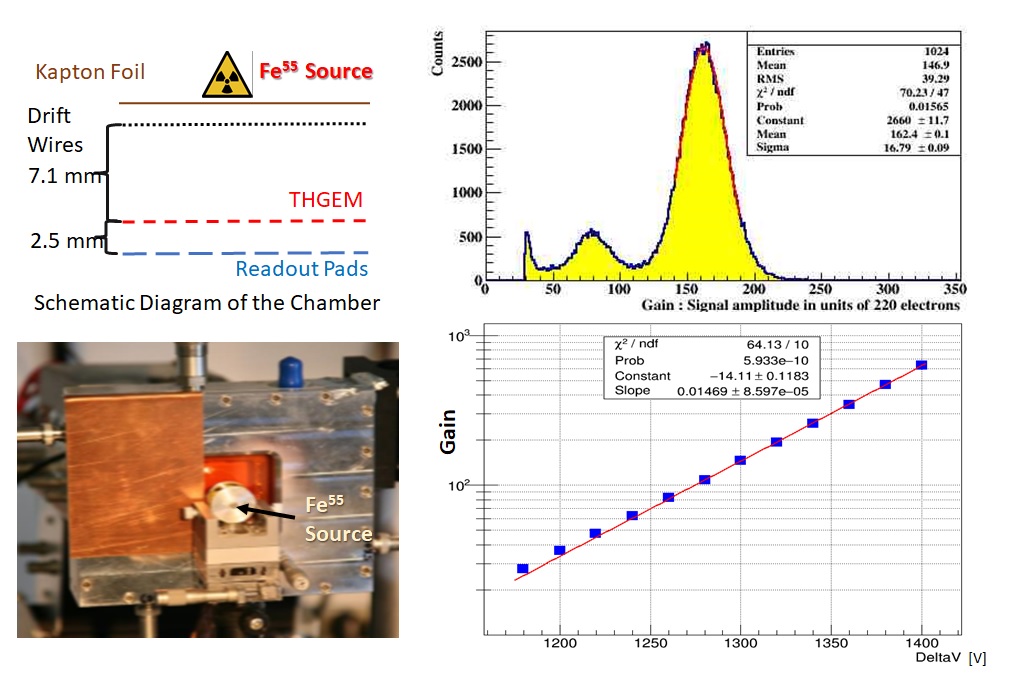}
\end{minipage}\hfill
\begin{minipage}[c]{0.98\textwidth}
        \caption{The schematic of our detector assembly (top-left panel).  The detector is illuminated with an ${}^{55}Fe$ X-ray source (bottom-left panel). A typical ${}^{55}Fe$ X-ray spectrum obtained in  $Ar:CO_{2}~=~70:30$ gas mixture with applied  voltages at drift, top and bottom of THGEM are 2520 V, 1720 V and 500 V respectively, anode is at ground (top-right panel). The bottom-right panel shows the gain dependence of THGEM versus the applied voltage.}
        \label{fig:Schematic_of_Detector_setup}
    \end{minipage}
\end{figure}

A detailed pre-characterizations of all THGEMs  had been performed at INFN, Trieste in order to have a comparative set of performance data as reference for the studies  after  coating with UV sensitive films.  The THGEMs have an active area of $30\times30~mm^{2}$. CsI and other UV sensitive material react with Cu. Therefore,  the THGEM Cu surface is coated with Ni-Au in order to isolate the Cu layer.  The  THGEMs used have hole diameter of  0.4~mm,  pitch of 0.8~mm, thickness of 470~$\mu$m including 35~$\mu$m of Ni-Au coated Cu on both faces. THGEMs with different rim sizes were used, where the rim is the clearance  ring around  the hole edge: $\le5\mu$m (no rim), $\sim10~\mu$m and $\sim20~\mu$m.
\par 

The detector assembly and some preliminary results of the  pre-characterization procedure are shown in figure \ref{fig:Schematic_of_Detector_setup}. In order to study the gain behavior of a THGEM,  we applied different voltages across it.  Before the voltage scan, the drift field (field above the THGEM) and the induction field (field below the THGEM) have been  optimized. This procedure has been repeated for each THGEM. After the coating,  we have studied the performance of the THGEMs in the same voltage configurations in order to do a comparative study and to see the effect of coating on each of the THGEMs.

 A typical ${}^{55}Fe$ X-ray spectrum obtained in  $Ar:CO_{2}~=~70:30$ gas  mixture is shown in   figure~\ref{fig:Schematic_of_Detector_setup}, top-right panel. The bottom right panel of figure \ref{fig:Schematic_of_Detector_setup} shows the gain dependence of THGEM versus the voltages applied across it.


\subsection{Coating procedure}

\begin{figure}
\begin{minipage}[c]{1\textwidth}
    \includegraphics[width=\textwidth]{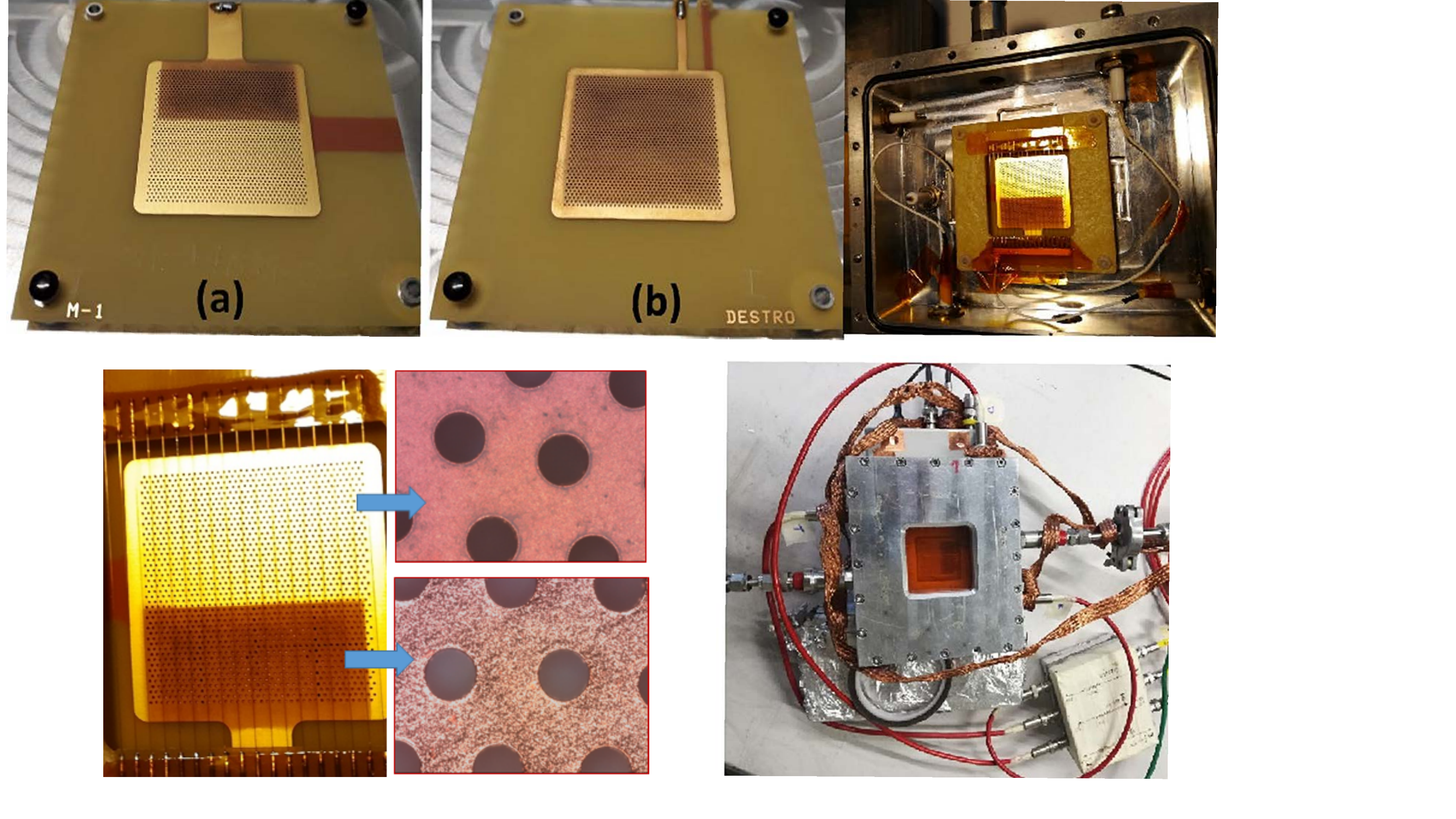}
\end{minipage}\hfill
\begin{minipage}[c]{0.98\textwidth}
        \caption{ THGEMs coated with ND powder covering (a) half and (b) fully the surface area.  Test chamber where the THGEMs are tested one at a time (top-right panel). A zoomed view of the both coated and uncoated parts of the THGEM  (bottom-left panel).  The test chamber after installation of a THGEM, with gas flow (bottom-right panel).}
        \label{fig:coating}
    \end{minipage}
\end{figure}

The reflective CsI film has been deposited by thermal evaporation technique at the Thin Film Lab (Dipartimento di Fisica-University of Bari).  The Microwave Plasma Enhanced Chemical Vapor Deposited (MWPECVD) diamonds are being used for producing UV photocathodes, which are more robust compared to CsI ~\cite{NIMA_447_2000_614, coating-I}. 
The hydrogenated poly- and nano-crystalline diamond films are deposited 
at high temperature, namely $800^{O}$C.
The high temperature prevents the use of THGEMs intolerant to this temperature scale. In order to get rid of these limitations, a novel and low-cost technique has been developed in Bari  ~\cite{coating-I, coating-II}. Nanodiamond powder with an average grain size of 250 nm produced  by Diamonds\&Tools srl have been used for the coating purpose. The H-ND is obtained by treating the as-received powder in $H_{2}$ microwave plasma for one hour at the MWPECVD Lab of CNR-ISTP in Bari. The ND and H-ND powder were separately dispersed in deionized water, sonicated for 30 minutes by a Bandelin Sonoplus HD2070 system and then sprayed  at $120^{0}$C on the THGEMs by coating half of them or fully their surface area (figure~{\ref{fig:coating}} (a) and (b)) with 400 or 800 spray pulses,  giving a granular layer formed by nanodiamond powder. 

The treated THGEM are listed in the following : 
\begin{itemize}
\item ~0 $\mu m$ rim  - ND half coated
\item ~0 $\mu m$ rim  - H-ND half coated
\item 10 $\mu m$ rim  - H-ND full coated 
\item 10 $\mu m$ rim  - CsI full coated
\item20 $\mu m$ rim   - ND half coated .

\end{itemize}

\section{Post-coating Characterization}


\begin{figure}
\includegraphics[width=.56\linewidth]{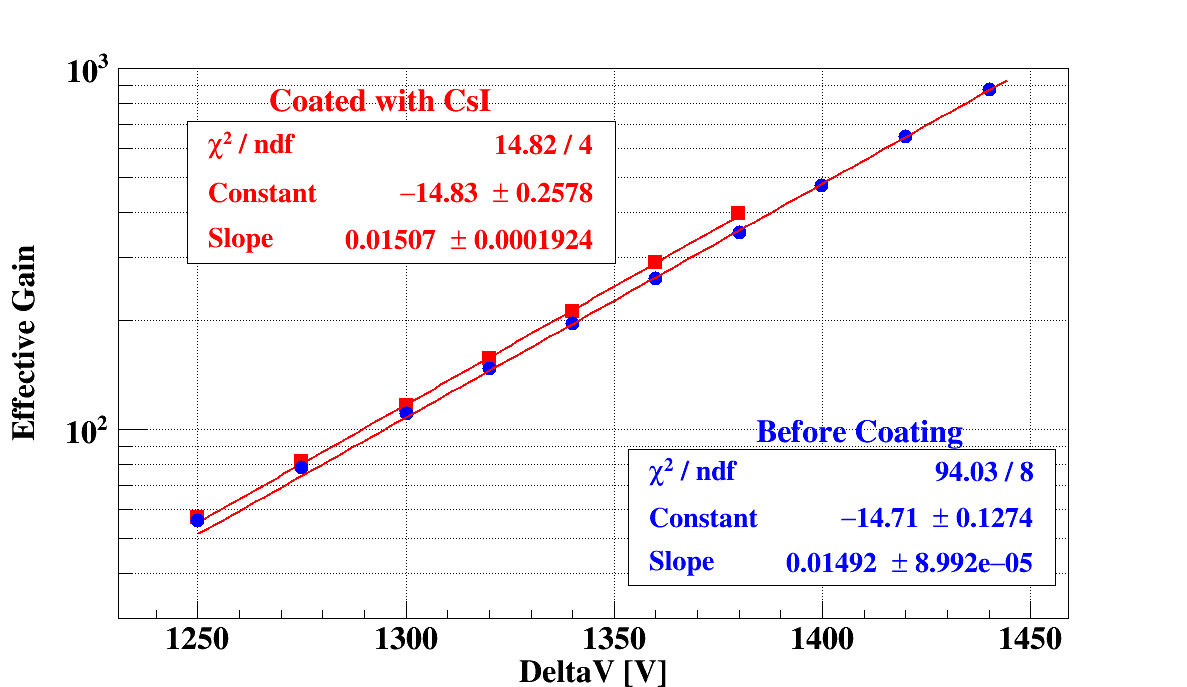}\hfill
\includegraphics[width=.44\linewidth]{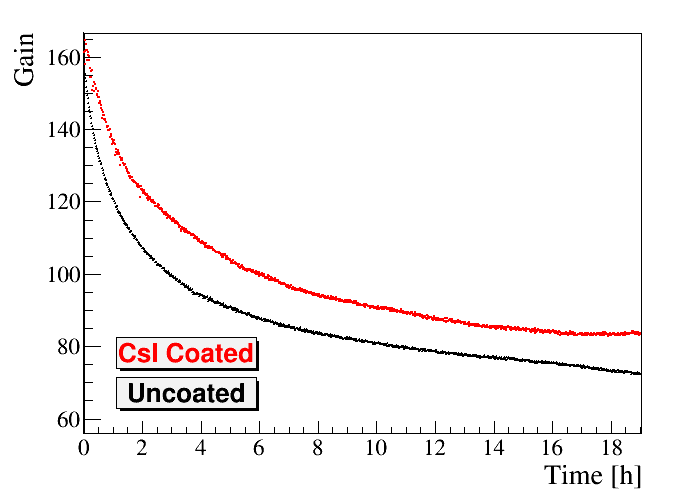}
\caption{The gains of the THGEM with 10~$\mu m$ rim measured before and after coating with CsI are compared. Gain versus applied voltage across a THGEM (left panel);  gain evolution versus time (right panel) .}
\label{fig:CsI_Comp}
\end{figure}


The characterization of THGEMs after coating with CsI, ND and H-ND, shows several interesting effects. The gain in the coated part is higher than the uncoated part. The factor of increment is however different on THGEMs with different rim size and with different coatings.  The THGEM with rim size $\sim10~\mu$m rim coated with a reflective CsI showed only a 20\% gain increment in comparison to the uncoated THGEM. The responses of the THGEM before and after coating are shown in figure \ref{fig:CsI_Comp}.

\begin{figure}
\includegraphics[width=.60\linewidth]{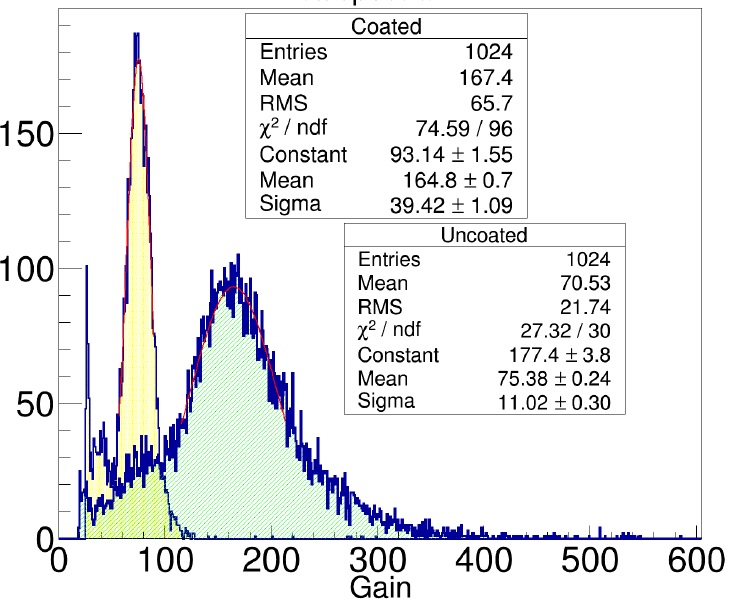}
\caption { ${}^{55}Fe$ X-ray spectra obtained with a 20~$\mu$m rim THGEM half-coated with ND powder.  The voltages applied  at drift, top and bottom of THGEM electrodes are 3510 V, 2110 V and 750 V respectively, while the anode is at ground. }
\label{fig:Gain_Jump}
\end{figure}

The gain variation in a THGEM with rim size $\sim20~\mu$m,  partially coated with ND, is shown figure \ref{fig:Gain_Jump}. 
In figure \ref{fig:Gain_Jump} we see a typical spectrum obtained illuminating with a  ${}^{55}Fe$   X-ray source for the both uncoated and ND coated parts at the same voltage. The gain in the coated part is a factor $\sim$ 2.2 larger than that in  the uncoated part.

\begin{figure}
\includegraphics[width=.5\linewidth]{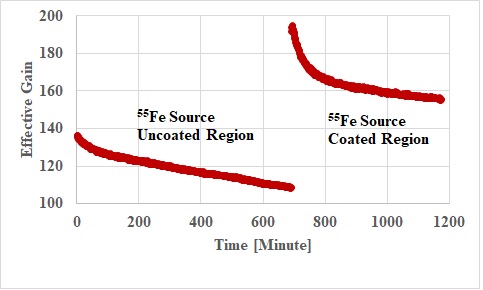}\hfill
\includegraphics[width=.5\linewidth]{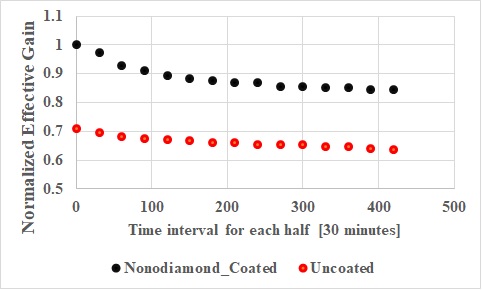}
\caption{Evolution versus time of the effective gain behavior of a THGEM with 0~$\mu$m rim, half-coated with ND. Gain versus time (left panel); the source is moved to the coated region at 700 minutes.   The same data normalized to the maxium gain measured in the coating region versus time, where t=0 is when the illumination of a region starts (right panel).}
\label{fig:Nor_Gain_LR}
\end{figure}

For a THGEM with no rim and half-coated with ND, the gain of the coated part is larger by a factor $\sim$1.4 (figure \ref{fig:Nor_Gain_LR}, left).  Concerning the gain evolution versus time,  the gain is maximum when the source illumination starts and then it decreases of ${\sim}$ 20\%. This effect is observed both for the uncoated and coated THGEM parts  (figure \ref{fig:Nor_Gain_LR}).


\par 
The THGEMs with H-ND coating, one with  0  $\mu m$ rim and a second one
with 10 $\mu m$ rim full, exhibit lower electrical stability compared to the THGEMs before coating. In fact, we have observed discharges in the THGEMs with fully coated H-ND at the nominal voltages.  These observations point to a potential resistivity issue related to the hydrogenated nanodiamond coating, to be confirmed by further studies. In spite of the instability, there are clear indication that, also in the case of H-ND coating, the THGEM gain is incremented when coated. 

\section{Conclusion}

Five THGEMs with and without rim as well as coated with different types of photosensitive layers have been characterized at INFN Trieste comparing the outcomes with a characterization exercise performed before coating. The coating of CsI, ND and H-ND  layers on THGEMs has been carried out at Bari, Italy.  Two interesting effects have been observed. Firstly, the increment of the gain in the coated part for the all photosensitive layers. Secondly, the lower electrical stability of the THGEMs coated with H-ND particles. The preliminary observation indicates that the THGEM with ND coating exhibit higher effective gain compared to those coated with CsI.

These results are very preliminary. A systematic study is required to explain the interesting observed effects. As a concluding remark, it can be stated that this newly grown technology is promising and can have wide applications in experimental particle physics and in many other scientific and industrial fields.



\section*{References}
\begin{thebibliography}{}

\bibitem{EIC}
A. Accardi et al., Electron-Ion Collider: The next QCD frontier, Eur. Phys. J. A52 (2016) 268.

\bibitem{windowless-RICH}
M. Blatnik et al., Performance of a Quintuple-GEM Based RICH Detector Prototype, IEEE NS 62 (2015) 3256.


\bibitem{PM18}
J. Agarwala et al., The MPGD-based photon detectors for the upgrade of COMPASS RICH-1 and beyond, Nucl. Instr. and Meth. A 936 (2019) 416.

\bibitem{NIMA_574_2007_28}
H.Hoedlmoser et al., Long term performance and ageing of CsI photocathodes for the ALICE/HMPID detector, Nuclear Inst. and Methods in Physics Research, A , 574 (2007) 28-38.

\bibitem{NDRep-1}
L. Velardi, A. Valentini, G. Cicala, Highly efficient and stable UV photocathode based on nanodiamond particles, Appl. Phys. Lett. 108 (2016) 083503–1-5.

\bibitem{NDreport}
A. Valentini, D. Melisi, G. De Pascali, G. Cicala, L. Velardi, A. Massaro, “High-efficiency nanodiamond-based ultraviolet photocathodes”, 30-03-2017 Patent n. WO 2017/051318 A9; International Patent n. PCT/IB2016/055616 of September 21, 2016; National Patent Italia - n. 102015000053374 del 21 Settembre 2015, Istituto Nazionale di Fisica Nucleare e Consiglio Nazionale delle Ricerche.

\bibitem{NIMA_502_2003_76}
F. Piuz, Ring imaging Cherenkov systems based on gaseous photo-detectors: trends and limits around particle accelerators, Nuclear Inst. and Methods in Physics Research, A , 502 (2003) 76-90.


\bibitem{NIMA_447_2000_614}
A.S. Tremsin, S. Ruvimov, O.H.W. Siegmund, Structural transformation of CsI thin film photocathodes under exposure to air and UV irradiation, Nuclear Inst. and Methods in Physics Research, A, 447 (2000) 614-618.


\bibitem{coating-I}
G. Cicala, A. Massaro, L. Velardi, G. S. Senesi, A. Valentini, Self-Assembled Pillar-Like Structures in Nanodiamond Layers by Pulsed Spray Technique, ACS Appl. Mater. Interfaces 6 (2014) 21101-21109.

\bibitem{coating-II}
L. Velardi, A. Valentini, G. Cicala, UV photocathodes based on nanodiamond particles: effect of carbon hybridization on the efficiency, Diam. Relat. Mater. 76 (2017) 1-8.

\end{thebibliography}
\end{document}